# Comparing victims of phishing and malware attacks
## Unraveling risk factors and possibilities for situational crime prevention

E.R. Leukfeldt

Faculty of Humanities and Law, Open University of the Netherlands

Cybersafety Reseach Group, NHL University of Applied Sciences & Police Academy of the Netherlands

e.r.leukfeldt@nhl.nl

*Abstract*—This paper compares the risk factors for becoming a victim of two types of phishing: high-tech phishing (using malicious software) and low-tech phishing (using e-mails and telephone calls). These risk factors are linked to possibilities for situational crime prevention. Data from a cybercrime victim survey in the Netherlands (n=10,316) is used. Based on routine activity theory, the multivariate analyses include thirty variables. There are many similarities between the two types of attacks. Both have almost the same crime script, for example. Furthermore, analyses show that both criminal groups are happy with almost any victim they can get and do not discriminate between the rich and poor. There are also differences between high- and low-tech victimization. First, almost none of the online visibility variables had any effect on the risk of becoming victim to low-tech attacks. Victims of high-tech attacks tell a different story. There is a clear connection between victimization and spending time in the online world. The second difference can be found within online accessibility. Again, none of the variables mattered when it came to the low-tech attacks. The malware analysis, however, shows that the victims' technical infrastructure does affect the risk of victimization. The findings show situational crime prevention has to be aimed at groups other than just the users themselves. Criminals are primarily interested in popular online places and the onus is on the owners of these virtual places to protect their visitors from getting infected.

I. INTRODUCTION

Everybody is at risk when it comes to phishing victimization.[1] This conclusion was drawn last year by Leukfeldt based upon an analysis of phishing victims. [1] The conclusion was both disappointing and intriguing. Disappointing because no one, clearly defined group could be pinpointed who are at increased risk of victimization. This makes target hardening a lot more difficult and limits specific prevention campaigns aimed at particular groups, e.g. those who engage in dangerous online activities. Intriguing because, based on routine activity theory, it was expected that, besides financial characteristics (value), online activities (visibility) and online accessibility in particular would play a role in determining the extent to which someone is a suitable target for these internet fraudsters.

Not one of the "digital" equivalents of visibility and accessibility, however, seemed to matter. The paper concluded that phishers apparently do not select their victims based on the digital trail victims leave when engaging in online activities, nor do they search for technical flaws in operating systems or browsers to get to users. A possible explanation was that it is the phisher's tactic to approach large groups of potential victims through, for example, spam. In this way, a fraudster can basically reach anybody with an internet connection and an e-mail account.

A recent study into phishing networks shows a phishing group which does indeed use such a method. [2] The group sends out as many e-mails as possible to Dutch citizens. In the e-mail it is stated that the security of their internet banking account needs to be updated or the account will be closed.[2] All the receiver has to do is click on the link in the e-mail to go to the bank's secure website. Of course, this is a fake bank website under the control of the criminals. Once someone tries to log in, this information goes to the criminals. From this moment on, the criminals have access to the bank accounts of the victims. Due to a layered security policy, however, the criminals are not yet able to transfer money from these accounts. This requires unique transaction authentication codes. The phisher then simply makes a telephone call to victims to try to obtain these codes. The caller poses as a bank employee and claims to need that code in order to complete the new security settings. Information from the victim's bank account is used to gain trust (e.g. the first (spam) e-mail about the new security, but also the type of account, registered names, last transactions, etc.).

The modus operandi outlined above supports the conclusion from the 2014 article that phishers use a trawling method. [1] Indeed, these phishers did not select their victims based on the digital trail they leave when engaging in online activities, nor did they seek for technical flaws in operating systems or browsers. Overall, technology (in this case the internet, e-mail and websites) is only used in the early stage of the crime. A major part of the crime script consists of using social engineering on victims. This explains why digital equivalents of the traditional elements of the routine activity theory show no effect on victimization.

---

[1] Phishing is the process aimed at finding out users' personal information by posing as a trusted authority using such digital means as e-mail.

[2] Or a similar message. The goal is to make the person click on the link in the e-mail.

The study into phishing networks [2], however, also shows that other types of networks exist. These criminals do rely heavily on technology to execute their attacks. They use malware[3], not only to steal user credentials but also to intercept transaction authentication methods or to alter transfers that victims have made. Indeed, these attacks depend more on technology than on social engineering. This raises the question whether the digital versions of the routine activity theory elements – value, inertia, visibility and accessibility (VIVA) – play a role within malware victimization.

A subsequent 2015 study into the applicability of routine activity theory to cybercrimes[4] indeed shows that digital equivalents of VIVA affect the risk of becoming a victim of malware [3]. This paper brings these two studies together, compares the risk factors, tries to explain the differences between the two types of offences and discusses which situational crime prevention methods are most suitable for the prevention of both low-tech and high-tech phishing.

Firstly, Section 2 briefly describes the expectations based upon routine activity theory. The data and methods are described in Section 3. Section 4 contains a description of the risk factors of malware victimization. Furthermore, the results of phishing and malware victimization are compared. The last section contains a conclusion and discussion about the differences and similarities between the two types of attacks, their victims and possibilities for situational crime prevention.

II. SUITABLE TARGETS: EXPECTATIONS BASED ON ROUTINE ACTIVITY THEORY[5]

This paper uses routine activity theory to find characteristics or behavior of victims which makes them more attractive to criminals. Routine activity theory claims that opportunity structures influence the prevalence of deviant behavior. [4] These structures are influenced by the combination of the presence of a motivated offender and a suitable target, and the absence of a capable guardian. There are four elements that determine the extent to which a victim appeals to a motivated offender: value, inertia, visibility and accessibility. Below, a brief description is given of each element and its digital equivalent, followed by the hypothesis.

---

[3] Malware is short for malicious software. Examples include viruses, worms, trojan horses and spyware.

[4] The focus of this study was not to identify risk factors, but to examine whether the classic elements (value, inertia, visibility and accessibility) of routine activity theory can still be used as a framework to study cybercrime.

[5] This paper compares the results of malware victimization with the findings of an earlier study. Hence, the same theoretical framework is used and the same variables will be used to test the hypotheses. This section, therefore, contains only a brief summary of the expectations based on the routine activity in the 2014 and 2015 study. [1][3]

The first element is value. Offenders are particularly interested in goals to which they assign value for whatever reason. In the case of phishing, the size of a bank account may be interesting as the profit potential is higher. Studies show households with higher incomes are more at risk of becoming victims of identity theft. [5][6]

- H1: Victims have significantly higher income and financial assets than non-victims.

The second element is inertia. Traditionally, inertia is described simply as the weight of the item. Small electronic goods, for example, are easier to steal than heavy cumbersome objects. In the case of information theft, this seldom applies (only with extremely large databases, at most). For this reason, the multivariate analysis excluded inertia.

Visibility is the third element. Visibility refers to the conspicuousness of objects criminals want, for example, expensive items in a living room that can be seen from the street. A number of studies on cybercrimes show online activities such as downloading or spending time in online chat boxes contribute to making someone a suitable target simply because they increase visibility. [7] [8] [9] [10] [11] [19] [22] [23] [24] In the case of phishing and malware, the question is which (and to what extent) online activities actually provide suitable targets. Respondents were asked to indicate what activities they conducted online (and how frequently). These were later classified in two categories: activities with a low or high level of online visibility. Activities within the first category are e-mail, targeted browsing and using direct messaging platforms like online messaging services and Skype. Activities in the second category are untargeted browsing, using online chat rooms, online gaming, online buying, actively using internet forums, active social networking and twittering.

- H2: Victims conduct significantly more highly visible online activities than non-victims.

The last element is accessibility. Accessibility is related to the construction of communities, placing goods in easily accessible locations and other features of everyday life that make it easy for offenders to come into contact with their target. Accessibility is also relevant in the online world. Users need software, such as operating systems and web browsers, to enter the online world. Motivated offenders abuse weaknesses in software to attack users. A relatively large group of motivated offenders is constantly trying to find new weaknesses in software and shares information on (potential) holes on forums (a number of studies describe these forums). [12][13][14[15][16] Indeed, so much is known about abusing weaknesses in popular software that users of popular software are more accessible to criminals than users of less popular software.

- H3: Victims use popular (commonly used) operating systems and web browsers significantly more often than non-victims.

There are also factors which make potential victims less accessible for attackers. There is, for example, anti-virus software which protects users against malware. Users become less accessible to a criminal when they take protective measures by installing and updating antivirus software. Furthermore, technical knowledge might also be a protective factor. The less users know about the software and equipment they use, the less they know about the risks they run. In addition, the victim's awareness of online risks may also determine their accessibility. Internet users who are aware of the risks they run online are better able to anticipate risk and are therefore less likely to become victims. [10][11]

- H4: Victims have significantly less up-to-date antivirus software than non-victims.
- H5: Victims have a significantly lower level of computer literacy than non-victims.
- H6: Victims have a significantly lower level of risk awareness than non-victims.

III. DATA AND METHODS

Both the phishing and malware analyses are based upon a secondary analysis of a Dutch cybercrime victim survey [17]. The representative sample included 10,314 respondents (response rate of 47%). The malware analysis involved 8,378 respondents (89%) who used the internet and answered questions about malware infections. Furthermore, data from the SSB (Social Statistical Database) of Statistics Netherlands was used in order to gain insight into the financial situation of respondents. See Arts and Hoogteijling for a detailed description [18].

The dependent variable – phishing victim – was coded dichotomously (1=victim, 0=no victim). Respondents were asked about phishing attacks that resulted in financial damage in the last twelve months. Of the 9,163 respondents who reported using the internet, 53 were victims of phishing (0.6%).

The dependent variable – malware victim – was also coded dichotomously (1=victim, 0=not a victim). Respondents were asked whether they had noted during the past twelve months that malware was present on their computer. Response options were: "yes", "no" and "I do not know". Of the 9,163 respondents who used the internet, 16.7 percent reported having malware on their computer(s).

The independent variables were: (1) sociodemographic traits such as gender, age, marital status, educational level (coded in eight categories from 'no education' to 'university education') and employment (12 hours per week or more); (2) additional financial data (personal income, household income, value of financial assets, amount of savings), added in collaboration with Statistics Netherlands. More about this in the next paragraph (3) frequency of online activities, rated by respondents on a four point scale; and finally (4), accessibility factors, such as computer skills (a composite variable of knowledge about the used operating system, internet connection, web browser and virus scanner) and risk awareness (a composite variable of ten propositions, such as: "I open attachments or files from unknown senders" and "I use different passwords for different accounts"). Other factors include type of operating system, web browser, and possessing an up-to-date virus scanner.

To gain insight into the financial situation of respondents, data from the SSB (Social Statistical Database) of Statistics Netherlands were linked to the dataset of Domenie et al (2013). The SSB is a database which is not publically accessible. It is only accessible to members of Dutch Universities. It consist of more than 40 linkable records about various subjects which are mutually matched. Data is provided to Statistics Netherlands by different government agencies like the police, IRS and social welfare agencies. A detailed explanation of the composition of the SSB files is given by Arts and Hoogteijling [18]. In order to link datasets with data from the SBB, a social security number or name and address is required. Statistics Netherlands then creates a so-called RIN number (an internal person identification number) to link data to individuals from different files. The researcher is then able to analyze data on a new file which is managed by Statistics Netherlands, and from which the social security number and / or name and address have been removed. Statistics Netherlands checks the output because any information which can be made publically may not be traceable to individual persons.;

Like any research, a number of comments can be made on the analyzed data. First of all, victim surveys rely on the answers of respondents. It might not be easy for respondents to distinguish phishing and malware attacks form each other. Furthermore, only online activities have been studied. Perhaps offline activities or psychological characteristics (also) play a role in phishing victimization. The dataset of the Dutch cybercrime survey [17] and additional data sets from Statistics Netherlands, however, did not provide such information. Other studies have found evidence that offline behavior does influence online victimization. [19] In addition, in previous research, a link with other theories was found, such as the self-control theory. Although self-control is rooted in offender studies, there is growing evidence that self-control also affects victimization [21][22]. In future studies, both offline behavior and such theories should be integrated in order to understand victimization better.

Finally, a comment on the usefulness of the presented multivariate analysis for (police) practice. The data of this article was collected in 2011. The (technical) developments within the cybercrime field take place with great speed. This also applies to a crime form such as phishing. To perform a successful phishing attack, criminals at all times use social engineering (persuading a user to do something which they normally would not do, for example, giving information about their internet banking account), whether or not combined with malware (an email with an infected attachment or a website where the visitor's computer is infected). Trend reports show an increase in the use of malware by phishers. Perhaps now or in the future, other risk factors play a role.

IV. SUITABLE TARGETS: RISK FACTORS COMPARED

Table 1 shows the results of both the multivariate analysis of phishing and malware victims. Even at first glance, it is clear there are differences between the two. Below the results are described for each of the elements of routine activity theory included in the analysis (value, visibility and accessibility).

*A. Value (hypothesis 1)*

Financial characteristics of respondents do not play a role in phishing victimization. None of the variables related to someone's wealth (e.g. income or savings) explains an increased or decreased risk of phishing victimization. Table 1 shows that most of the financial characteristics do not play a role within malware victimization either. One variable, however, is related to the risk of becoming a victim: the level of someone's personal income. The effect, however, is not in the expected direction: the lower the personal income, the higher the probability of becoming a victim. Why a lower income increases the risk of malware victimization is unclear. This should be the subject of further research.

The analysis clearly shows that phishing attacks (both low-tech and high-tech) are not only aimed at potential victims with high levels of funds in their bank account. Again, there seems to be no evidence for so-called spear-phishing attacks on specific targets with lots of money. Apparently, these criminals are happy with anything they can get and do not discriminate between the rich and poor.

[Table 1]

*B. Visibility (hypothesis 2)*

The visibility variables are divided into activities with high and low visibility. Regarding phishing, none of the activities with high visibility are related to victimization. Some of these activities, however, do appear to increase the risk of malware victimization.

Perhaps the most obvious one is downloading. The more one downloads, the greater the chance of getting a malware infection. Other studies already showed downloading can be a risky activity. One of the ways to infect computers is by letting the user open a file which contaminates the computer. All the criminal has to do is make sure the user opens the file. This can be done by disguising malware as a nice picture or including it in the download of (what the users thinks is) the latest movie or software. Another option is that users are lured to websites which offer the latest downloads of whatever. As soon as the user enters the site, malware is transmitted.

This last explanation might also be true for online games. This activity also increases the risk of victimization. Again, if users are actively looking for new games to play online, they might end up at infected sites which pretend to offer the latest online games.

Buying goods online is another activity with high visibility which increases risk of malware victimization. A possible explanation can be found in the goal of malware: intercepting user credentials to commit fraud. People who buy more often online are also likely to pay more often online, or at least have to share their financial credentials more often online with the company or persons with whom the business is done. Attackers simply have more opportunities to intercept credentials.

This last explanation is in line with the other type of online activities which enlarge chances of malware victimization: frequency of internet use, targeted and untargeted browsing. The more users carry out these common activities, the greater the chance of becoming a malware victim. When it comes to phishing victimization, only targeted browsing seemed to be risk enhancing. This was an unexpected finding with the possible explanation that not phishing, but malware infections caused this effect. Users can get contaminated by visiting infected websites. These website are not by definition located in the dark alleys of the internet. In recent years, for example, many popular, legitimate websites had advertisements which were contaminated and infected visitors to that website. The conclusion was that "Internet users who visit all kinds of (legitimate) sites are at greater risk." This explanation is clearly being confirmed by the current analysis: being online, surfing the web (both targeted and untargeted) is a risk-enhancing activity for malware infections.

*C. Accessibility (hypothesis 3-6)*

Again, when it comes to phishing victimization, none of the factors in relation to accessibility have a risk-increasing effect. This applies to both online accessibility (use of popular operating systems and web browsers) and protective measures to reduce accessibility (up-to-date virus scanner, computer knowledge and risk perception). Apparently, in the case of phishing, technical accessibility is not of major importance. This corresponds with the modus operandi of phishing networks covered in the introduction: technology is only used in the early stage of the crime; the major part of the crime script consists of using social engineering on victims. No large-scale, high-tech attacks, but tailored low-tech attacks.

But what about malware? Table 1 shows that the use of a certain operating system and web browser is risk increasing. Users with a Windows operating system have an increased risk of malware victimization. This may be because this is a widely used operating system. It is simply more profitable for criminals to write malware for this system because there are more potential victims. It also appears that users have an increased risk of victimization with the Firefox browser. At the time of data collection for this study this was the most popular browser after Internet Explorer.[6]

Another technical aspect without any effect on both phishing and malware victimization is the use of a virus scanner. The conclusion made in 2014 article [1] remains intact: a virus scanner cannot guard against mail that persuades users to provide personal information (low-tech attack). It might protect against some forms of malware (high-tech attack), but only to known variants. New variants are not detected (yet). Virus scanners also do not protect against criminals who abuse zero-day exploits (a flaw in software for which there is no patch at a certain moment).

## V. CONCLUSION AND DISCUSSION

There are many similarities between (low-tech) phishing attacks and (high-tech) malware attacks. Both type of attacks have, for example, almost the same crime script. Furthermore, both groups of criminals are happy with anything they can get and do not discriminate between the rich and poor.

There are also differences. Firstly, regarding the effect of online visibility, almost none of the variables had any effect on phishing victimization. The victims of malware, however, tell a different story. There is a clear connection between spending time in the online world and victimization. There are two types of activities which ensure a higher chance of becoming a victim, namely downloading and online gaming. The second difference can be found within online accessibility. Again, none of the variables mattered when it came to phishing. The malware analysis, however, shows that users with certain (popular and widely used) operating systems and web browsers do have an increased risk of victimization.

What do these findings mean for situational crime prevention? The conclusion regarding phishing victimization was that everybody is at risk. Opportunities for prevention campaigns aimed at a specific target group or dangerous online activities are limited. Situational crime prevention will have to come from a different angle, for example, the banks taking on the role of capable guardian.

At first glance, the malware analysis shows more opportunities for prevention campaigns aimed at specific groups because there are a number of variables which are risk enhancing. When we take a closer look, however, the analysis shows that obvious – and relatively easy to adjust – protective variables, like using a virus scanner, computer knowledge and online risk perception, have no effect on victimization.

The risk seems to come from two different angles. The first one is spending more time online. Just visiting all sorts of (legitimate) websites is risk enhancing. More research on exactly how and where victims got infected has to be done, but in the light of routine activity theory it is expected that criminals aim their attacks at popular online places (like the recent infections of advertisements on popular websites). The more visitors a website has, the larger the amount of potential victims. The second group who is at risk are people who are looking for downloads or online games. Again, the law of large number applies: popular downloads and games can be used to trick as many potential victims as possible to visit an infected website, or to download an infected file.

These findings underline the conclusions of the 2014 article [1]: situational crime prevention has to be aimed at other groups than just the users themselves. The malware analysis shows there is an important role for owners of (popular) websites. Indeed, criminals are interested in these online places and the owners of these places have to protect their visitors from getting infected.

Furthermore, the current analysis points out the way for further (in-depth) studies. Firstly, classic methods like interviews with victims might shed light on some of the findings. What type of websites are the victims actually visiting? Do they know on which site or because of which download they got infected? Why did they visit that particular website? Did they have a feeling something was wrong? Subsequently, more pioneering methods like analyzing logs of user activity from Internet Service Providers or examining computers and browsers of users their selves by researchers might provide better insight into actual activities of victims. In addition to this, criminal networks need to be analyzed more systematically. Leukfeldt's study [2] showed that criminal networks with the same crime script may have major differences in their nature (low-tech vs. high-tech). As the current analysis shows, this might affect the type of victim they target. Insight into what other types of networks there are (origin and growth, type and role of offenders, criminal capabilities, etc.) might give us insight in how these networks actually select their victims.

ACKNOWLEDGMENTS

This study is part of the Dutch Research Program Safety and Security of Online Banking. This program is funded by the Dutch banking sector, represented by the Dutch Banking Association (NVB), the Police Academy and the Dutch National Police. Executive organizations are the Open University of the Netherlands, the Dutch Police Academy and NHL University of Applies Sciences.

The original dataset is obtained from the study of Domenie et al (2013). This research was funded by the Nation Unit and the Cybercrime Program of the Dutch police. Executive organizations were the Open University of the Netherlands, NHL University of Applies Sciences, Dutch Police Academy, Leiden University and the Statistics Netherlands.

---

[6] http://gs.statcounter.com/#browser-NL-monthly-201101-201106

*Table 1: Logistic regression of phishing and malware victims*

|  |  | Phishing [1] |  | Malware [3] |  |
|---|---|---|---|---|---|
|  |  | B | S.E. | B | S.E. |
| Constant |  | -10,685 | 3,176 | -6.296*** | .591 |
| Value |  |  |  |  |  |
|  | Personal income | .000 | .006 | -.005** | .001 |
|  | Household income | -.002 | .007 | .002 | .001 |
|  | Financial assets | .000 | .001 | .000 | .000 |
|  | Financial possessions | .000 | .001 | .000 | .000 |
|  | Savings | .000 | .000 | .000 | .000 |
| Online visibility |  |  |  |  |  |
|  | Frequency of internet use | .231 | .498 | .233** | .091 |
| Online activities with low visibility |  |  |  |  |  |
|  | Targeted browsing | .519* | .253 | .156** | .048 |
|  | Direct communication: e-mail | .197 | .199 | -.049 | .041 |
|  | Direct communication: MSN. Skype | .066 | .151 | -.020 | .034 |
| Online activities with high visibility |  |  |  |  |  |
|  | Chatting in chat boxes | .187 | .208 | -.015 | .053 |
|  | Online gaming | -.006 | .167 | .084* | .034 |
|  | Active on online forums | -.132 | .261 | .142 | .051 |
|  | Active on social network sites | .135 | .153 | -.015 | .034 |
|  | Twitter | -.024 | .229 | .014 | .050 |
|  | Downloading | -.331 | .331 | .078* | .033 |
|  | Untargeted browsing | .016 | .144 | .139*** | .033 |
|  | Buying online | .159 | .369 | .387*** | .084 |
| Digital accessibility |  |  |  |  |  |
|  | OS: Windows | -.122 | .649 | .912*** | .184 |
|  | Browser: Internet Explorer | -.115 | .383 | .125 | .085 |
|  | Browser: Google Chrome | -.072 | .360 | .089 | .076 |
|  | Browser: Firefox | .0353 | .356 | .173* | .082 |
|  | Browser: Opera | .491 | 1.049 | .289 | .265 |
|  | Browser: Safari | .128 | .636 | -.362 | .188 |
|  | No virus scanner | .255 | .562 | -.111 | .154 |
|  | Computer knowledge | .643 | .504 | .167 | .090 |
|  | Online risk perception | -.145 | .266 | -.007 | .059 |
| Nagelkerke $R^2$ |  | 5.0 |  | 8.4 |  |
| N |  | 8,379 |  | 8,378 |  |

* p<0,01; ** p<0,001